\documentstyle[prl,twocolumn,aps]{revtex}
\begin{document}
\draft
\title{Extraction of Step-Repulsion Strengths from Terrace Width
Distributions: Statistical and Analytic Considerations}
\vspace{-0.3cm}
\author{Howard L.\ Richards, Saul~D.\ Cohen, and
        T.~L.\
Einstein\thanks{Corresponding author:
        {\tt einstein@physics.umd.edu}, fax: (+1) 301-314-9465
\newline \newline \newline \setlength{\baselineskip}{8pt}
}}

\address{\setlength{\baselineskip}{16pt} Department of Physics,
University of Maryland, College Park, MD 20742-4111\\}
\vspace{-0.3cm}
\author{M.~Giesen}

\address{\setlength{\baselineskip}{16pt} Institut f\"{u}r
Grenzfl\"{a}chenforschung und Vakuumphysik, Forschungszentrum
J\"{u}lich,\\ D-52425 J\"{u}lich, Germany}

\date{April 12, 2000}
\maketitle
\begin{abstract}
Recently it has been recognized that the so-called generalized Wigner
distribution may provide at least as good a description of terrace
width distributions (TWDs) on vicinal surfaces as the standard
Gaussian fit and is particularly applicable for weak repulsions
between steps, where the latter fails.  Subsequent applications to
vicinal copper surfaces at various temperatures confirmed the
serviceability of the new analysis procedure but raised some
theoretical questions. Here we address these issues using
analytical, numerical, and statistical methods. We propose an
extension of the generalized Wigner distribution to a two-parameter
fit that allows the terrace widths to be scaled by an optimal
effective mean width.  We discuss quantitatively the approach of a
Wigner distribution to a Gaussian form for strong repulsions,
how errors in normalization or mean affect the deduced interaction,
and how optimally to extract the interaction
from the variance and mean of the TWD.  We show that
correlations reduce by two orders of magnitude the number of
{\em independent} measurements in a typical STM image.
We also discuss the effect of the discreteness (``quantization") of
terrace widths, finding that for high misorientation (small mean
width) the standard continuum analysis gives faulty estimates of step
interactions.
\end{abstract}

\section{Introduction}

During the last decade a number of
researchers have used atomic-scale microscopy to make quantitative
experimental measurements of the terrace width distribution (TWD)
of vicinal surfaces.  To understand the data --- and,
especially, to extract the strength of the interaction
between the steps --- they have fit the TWDs with Gaussians
(or in cases of no apparent energetic repulsion, with
free-fermion distributions).  Recently there has been a
significant improvement in the theoretical understanding of
interacting steps on vicinal surfaces: as an example of a
fluctuation phenomenon, they should be described by certain
universal features related to random-matrix theory.  In particular,
the TWD should be well describable in terms of a generalized form
of the distribution surmised by Wigner to describe some special
cases of interactions\cite{EP99}.

In a recent paper\cite{GE}, hereafter GE, terrace width distributions
(TWDs) of various vicinal copper surfaces were analyzed using both the
traditional Gaussian approach and the generalized Wigner surmise.  Many
conclusions were noted in passing about the relative merits and
sensitivities of these two approaches.
The goal of this paper is to provide supporting details together
with new results and approaches that should aid in the interpretation
and analysis of experimental TWDs. We explore the relationship
between the Wigner form of TWDs and the Gaussian.  We discuss several
statistical considerations that should be taken into account. The
many issues treated by this paper arose during the course of
analyzing experimental data in GE.

The organization of this paper is as follows.
Sec.~\ref{sec-series} reviews the TWD derived from the generalized
Wigner surmise and presents some practical new approximations derived
from series expansions.
In particular, we provide what we believe is the best simple
expression [Eq.~(\ref{e:as})] to deduce the step-step repulsion
strength from the variance of the TWD.
Sec.~\ref{sec-gausswigner}
deals with the approach of the generalized Wigner distribution
to the form of the Gaussian for strong step-step repulsions.
While this behavior had been recognized earlier,
we now characterize it quantitatively.
In section~\ref{sec-normmean}, we contend with a recurring
theme in GE: the error generated by uncertainty in
the mean of the distribution.
Experimentalists had the belief that Gaussian fits of data are more
forgiving  of such errors than are Wigner fits.  We study this notion
quantitatively by checking for both distributions the effect of
perturbations in normalization and in mean
by fitting deliberately misnormed or displaced
data.  The results of arguably greatest interest to experimentalists are
in section~\ref{sec-wigner2param}.
We describe a extension of our proposed analysis scheme for TWDs
for which the first moment of the data does not conform well to the
apparent mean.  We propose treating the generalized Wigner distribution
as a {\em two-parameter function}: in addition to the exponent
$\varrho$, the value of the effective mean (which scales the terrace
widths; cf.\ Sec.~\ref{sec-series}) is adjusted simultaneously in the
non-linear least-squares fit. This procedure makes little
difference for the ``good'' data reported in GE, but can have
significant effect on ``poorer'' data glossed over in that paper.
We present both graphical illustrations and thorough tabulations for
the extensive data for vicinal copper discussed in GE.  We also apply the
Wigner distribution to recently published data for vicinal Pt(110).
Sections~\ref{sec-discrete} and
\ref{sec-sample} offer a pair of warnings regarding how the discreteness
of the terrace widths and the limited size of the sample, respectively,
can confound the analysis.  In the former case, for the range of
interaction strengths found in physical systems, discreteness becomes
problematic for high misorientations, when the mean terrace width drops
to just a few lattice spacings. In the latter case, we observe that
statistical fluctuations due to the typical size might well account for
some of the data sets labeled as ``poor," rather than some system
contaminant or measurement flaw. A conclusion summarizes the current state
of our understanding.

\section{Generalized Wigner Surmise: Recap of Key Formulas and New Results
from
Series Expansion}
\label{sec-series}

As has been discussed extensively before\cite{GE,EP99},
a new idea from random-matrix theory\cite{MehtaRanMat,Guhr98}
is that fluctuations should exhibit certain universal behavior.
According to the so-called Wigner surmise, the distribution of
fluctuations can be approximated by\cite{EP99}
\begin{equation}
  \label{e:Wigner}
   P_\varrho(s) = a_\varrho s^\varrho \exp \left(-b_\varrho s^2\right) \,
,
\end{equation}
where $s\equiv\ell/\langle\ell\rangle$, $\ell$ being the terrace width,
and the constants
$b_\varrho$ and
$a_\varrho$ are given by
\begin{eqnarray}
    b_\varrho & = & \left[\frac{\Gamma \left(\frac{\varrho+2}{2}\right)}
      {\Gamma \left(\frac{\varrho+1}{2}\right)}\right]^2 \nonumber \\
 & \approx &
\left(\frac{\varrho+1}{2}\right)
      \left[1 - \frac{1}{2}(\varrho+1)^{-1}
             + \frac{1}{8}(\varrho+1)^{-2}\right] \nonumber \\ & = &
\frac{\varrho}{2} +\frac{1}{4}
             + \frac{1}{16(\varrho+1)}
\label{e:br}
\end{eqnarray}
and
\begin{equation}
    a_\varrho = \frac{2b_\varrho^{(\varrho + 1)/2}}
	          {\Gamma \left(\frac{\varrho+1}{2}\right)}
           = \frac{2\left[\Gamma
\left(\frac{\varrho+2}{2}\right)\right]^{\varrho+1} }
                  { \left[\Gamma
\left(\frac{\varrho+1}{2}\right)\right]^{\varrho+2} } \, ,
\label{e:abr}
\end{equation}
respectively.  For brevity, we refer hereafter to this set of formulas as the
CGWD (continuum generalized Wigner distribution).
The CGWD can be derived in a more transparent fashion from
a mean-field approximation\cite{beyond}.

The approximate result in Eq.~(\ref{e:br}), derived in
Appendix A by asymptotic expansion, is new.
It is consistent with Eq.~(9) of GE in the neighborhood of
$\varrho\! \approx \! 4$; it is within 0.2\%
of the exact $b_\varrho$ as calculated using gamma functions
at $\varrho = 2$ and is within 0.05\%
of $b_\varrho$ by $\varrho = 4$.

Experimentally, a TWD is typically characterized by
its variance $\sigma^2$.
In principle $\sigma^2$ might be determined directly from the second
moment of the TWD, but there is concern that this approach does not
adequately minimize noise in the data, an issue we shall revisit in
Sec.~\ref{sec-sample}.  Thus, in practice, TWDs are fit to smooth
functions; Gaussians are typically chosen, not just for their simplicity
but because their use can be justified readily in the limit of strong
elastic
repulsion between steps.  The variance of the TWD is then approximated by
the
variance
$\sigma^2_G$ of the fitted  Gaussian.  We argue here and in GE that the
CGWD given in Eq.~(1) is scarcely more
complicated than a Gaussian but provides a better accounting of the
variance. For strong step repulsions, the variance of the fitted Gaussian
is  usually
not very different from the variance
$\sigma^2_W$ of a CGWD, as is
discussed more quantitatively in Sec.~\ref{sec-gausswigner}. For weak
repulsions, however, it is well known that the TWD becomes too skewed to
allow a satisfactory fit to a Gaussian.  Experimentalists finding
themselves in this predicament have been stymied on how to proceed
quantitatively \cite{SBV,HFNH}. Significantly, a Gaussian fit to a TWD
with nonnegligible skewness cannot even be expected to have the correct
mean;
the consequences of this fact are dealt with in much of the remainder of
this
paper.

For the CGWD, the variance can be
expressed simply in terms of $b_\varrho$.   We can use Eq.~(\ref{e:br})
to obtain

\newcounter{eqsave}%
\setcounter{eqsave}{\value{equation}}%
\stepcounter{eqsave}%
\setcounter{equation}{0}%
\renewcommand{\theequation}{\mbox{\arabic{eqsave}\alph{equation}}}%
\begin{eqnarray}
\label{e:sra}
  \sigma^2_W  & =  & \frac{\varrho + 1}{2b_\varrho} - 1  \\
            & \approx & \frac{1}{2}(\varrho + 1)^{-1}
                      + \frac{1}{8}(\varrho + 1)^{-2}
\label{e:sr}
\end{eqnarray}
\setcounter{equation}{\value{eqsave}}%
\renewcommand{\theequation}{\arabic{equation}}%

\noindent for large values of
$\varrho$ (e.g.\ $\sigma^2$ is overestimated by about 0.5\%
at $\varrho = 4$
but just 0.1\%
at $\varrho = 10$).

The usual goal in an experiment is to extract the magnitude $A$
of the elastic repulsion between steps, perpendicular to the
step direction, given by $A/\ell^2$.
All standard analysis procedures make a continuum approximation in
the direction along the steps (perpendicular to the ``upstairs"
direction); thereafter, $A$ appears only in the form of a dimensionless
interaction strength
$\tilde{A} \equiv A\tilde{\beta}(k_{\rm B}T)^{-2}$,
where $\tilde{\beta}$ is the step stiffness.
In this conceptualization $\varrho$ is related
to $\tilde{A}$ by the equation
\begin{equation}
   \tilde{A}_{\rm W} = \varrho (\varrho -2)/4 \,
\label{e:ar}
\end{equation}
\noindent which follows from mapping this problem onto the
Sutherland Hamiltonian\cite{s71}. The subscript $W$
provides a reminder that this estimate of $\tilde{A}$ is based on the
CGWD. Eq.~(\ref{e:sr}) can be solved for $\varrho$,
which in turn can be inserted into Eq.\ (\ref{e:ar}) to
provide a good estimate for $\tilde{A}_{\rm W}$.  However,
a much better estimate of $\tilde{A}_{\rm W}$---visually
indistinguishable from the exact value on a
standard-resolution graph---comes from performing a reversion of
series of  a higher-order version of Eq.~(\ref{e:sr}) to
yield $\varrho$ as a function of $\sigma^2$,
\begin{equation}
  \label{eq:rho_of_sig2}
  \varrho \approx  \frac{1}{2}\left(\sigma^2\right)^{-1}
           \left\{ 1
             - \frac{3}{2} (\sigma^2)
             - \frac{3}{4} (\sigma^2)^2
             + \frac{7}{24}(\sigma^2)^3 \right\}\, ,
\end{equation}
and then inserting this result into Eq.~(\ref{e:ar}):
\begin{equation}
  \label{e:as}
  \tilde{A}_W
    \approx  \frac{ 1}{16} \left[(\sigma^2)^{-2}
          - 7 (\sigma^2)^{-1}
          + \frac{27}{4}
          + \frac{35}{6} \sigma^2\right]  \, .
\end{equation}
\noindent Eq.~(\ref{e:as}) should prove quite useful in analyzing data,
since it provides an excellent value for $\tilde{A}$ as a function of the
variance of the TWD, assuming the validity of the CGWD description.
We caution that {\em all four terms must be kept} in order to obtain a good
estimate of $\tilde{A}$ from Eq.~(\ref{e:as}).  We also warn that, as
discussed
in Sec.~\ref{sec-discrete}, the effects of discreteness may lead to
inconsistencies with this estimate for highly misoriented vicinal
surfaces.

\section{Gaussian Fits of the Generalized Wigner Distribution}
\label{sec-gausswigner}

 A characteristic feature of the CGWD is
that as $\varrho$ becomes larger, the curve can be better approximated
by a Gaussian.  This feature should be expected, since it is accepted that
TWDs for strong repulsions are well described by Gaussians.  We
quantitatively assessed the degree of agreement.
 One measure is the $\chi^2$  of a fit of the CGWD to
a Gaussian form (with the three parameters---peak position, prefactor,
and standard deviation---as adjustable parameters).  We find that
this measure of the fit improves exponentially with increasing
$\varrho$.  (Specifically, $0.012144 \exp(-0.5249 \varrho)$ provides a
close upper bound of $\chi^2$ for $\varrho > 1$.)  A second and more
useful measure is the relative difference of the standard deviation
$\sigma_G$ of the fitted Gaussian from the actual standard deviation
$\sigma_W$ of the CGWD, given by the square
root of the second moment of the CGWD about its mean
of unity.  Using Eq.\ (\ref{e:br}) we find
that this relative difference is well described  by the formula
\begin{equation}
  	\frac{\sigma_G}{\sigma_W} -1 \equiv
	\frac{\sigma_G}{\sqrt{\mu_2^{\prime}-1}} - 1 \approx
	\frac{0.0568}{\varrho} - \frac{0.0138}{\varrho^2} \, ,
\label{e:method2}
\end{equation}

\noindent where the expression for the second moment of the
TWD with respect to the origin, $\mu_2^{\prime}$, is given explicitly as
Eq.\
(11) of GE or Eq.\ (8) of EP.  Thus, at the calibration point for
repulsive
interactions  ($\varrho =4$, for which an exact solution exists) the
agreement
is  around 1\%,
and improves monotonically with increasing
$\varrho$.  For  this range ($\varrho \geq 4$) differences between estimates of
$\tilde{A}$ obtained from CGWD and the various Gaussian fit
methods are predominantly  due to different philosophies of
extracting $\tilde{A}$ from
$\sigma$ rather than from differences in the fitting methods.

As discussed at length in EP and GE, there are
several distinct theories for extracting the dimensionless interaction
strength $\tilde{A}$ from $\sigma_G$.  Monte Carlo calculations \cite{CRE}
indicate that  the CGWD provides an excellent estimate of
$\tilde{A}$ over the  range of physical values of this repulsion, as
well as for stronger  values.  Thus, as remarked at the end of the
previous
section, it is the wisest strategy to use  Eq.~(\ref{e:as}) to estimate
$\tilde{A}$ from
$\sigma$ deduced from the TWD  rather than to use the predictions of one
of
the Gaussian approximations discussed in Table~1 of GE.

\section{Effects of Perturbed Normalization or Mean}
\label{sec-normmean}

The CGWD is a normalized TWD with unit mean.
In GE, the mean was determined straightforwardly from
the first moment.  The independent variable (the terrace width)
was then scaled by this value, and the distribution normalized.
In the course of analyzing TWDs, it became obvious
that the normalization of the data sets
by total area (that is, the zeroth moment)
and first moment provides qualitative agreement with
the CGWD
--- that is, the ``best fit'' CGWD produces a skew distribution
that roughly matches the TWD --- but it does not match closely
enough to reproduce the correct peak position.
In order to motivate the more satisfactory treatment of experimental
TWDs in section~\ref{sec-wigner2param}, in
this section we discuss the effects of perturbations of
the mean step separation and of the normalization on
the variables important for extracting interaction strengths
($\sigma$ for a Gaussian fit and $\varrho$ for a Wigner fit).
Such perturbations might arise in experimental
data either due to statistical fluctuations or due to physical causes,
such as perturbations of the step-step interaction potential
$A/\ell^2$ or an incomplete equilibration of the vicinal
surface.

{}To this end, we created an ideal data set
by sampling the appropriate distributions at regular intervals.
This ideal set was then perturbed by various factors not exceeding 15\%,
either by shifting the mean or by scaling each point to increase the
area under the curve.
These perturbed sets were then fit as in GE,
by normalized fitting functions with unit mean.  Since the true
value of $\varrho$ or $\sigma$ for our ideal data set is known,
it is simple to determine the error due to the perturbations.
In some cases, the errors behave in complicated ways.

In the equations, $\Delta \sigma$ is the
fitted value of $\sigma$ minus the known value of $\sigma$ (and similarly
for
$\Delta \varrho$); $\Delta \mu_0$ (or $\Delta \mu_1$) indicates how much
the area under (or the first
moment of) the constructed curve exceeds the ``proper value."
(Moments about the origin are defined in Eq. (10) of GE.  Here for
convenience---since we are interested only in differences---we neglect the
primes.  The effects of errors in normalization can be described rather
simply. The fitted [normalized] curve becomes narrower as the area under
the raw curve increases.  For a Gaussian, the fractional change
in $\sigma$ is approximately linear in the fractional error of the
integrated
TWD, with a prefactor about 2/3:

\begin{equation}
\Delta\sigma/\sigma|_{\sigma=0.30}=-0.68\Delta\mu_0
+0.81(\Delta\mu_0)^2
\label{e:ds}
\end{equation}

The coefficients in Eq.\ (\ref{e:ds}) are insensitive
to the value of $\sigma$: if the standard deviation of the raw curve
is reduced from 0.30 to 0.20, the linear coefficient is unchanged,
while the quadratic coefficient is reduced slightly to 0.80.
 For the CGWD, the fit is even more nearly linear:

\begin{equation}
\Delta\varrho/\varrho|_{\varrho=4.0}=1.38\Delta\mu_0
\label{e:dr}
\end{equation}

Again increased area leads to an effectively
sharper distribution.  The linear coefficient is nearly double
that in Eq.\ (\ref{e:dr}), as one might expect from Eq. (13) of GE.
This  coefficient again is insensitive to the value of $\varrho$ of the
raw  distribution: for $\varrho$ = 7.0 it dips slightly to 1.37.

 Errors in the mean of the distribution create errors in the
fit that are not so easy to describe.  The changes in the fitted
parameters are quadratic rather than linear in $\Delta \mu_1$, and
the coefficients  depend strongly on the value of $\sigma$ or $\varrho$
of the raw distribution.

For Gaussians, we find that the following expression provides
a good approximation for standard deviations between 0.2
and 0.4 (corresponding to 1.5 $< \varrho <$ 9):

\begin{equation}
\Delta\sigma/\sigma 
\approx (1/2)(\Delta\mu_1/\sigma)^2 .
\label{e:Ds}
\end{equation}
\noindent Appendix B provides an analytic derivation of this approximation
as the leading-order term in an expansion of the appropriate Gaussian
integral.  Eq.~(\ref{e:Ds}) can also be generated from straightforward fitting
of numerical data.\footnote{In the process, one can generate the
result $\Delta\sigma/\sigma \approx 0.486\sigma^{-2.05}(\Delta\mu_1)^2 $,
which is numerically superior to Eq.~(\ref{e:Ds}) but does not satisfy proper
dimensional behavior.}

Thus, as might be expected since the Gaussian is symmetric about
its peak, the error is insensitive to the sign of the error in
the mean of the raw distribution.  The fitted distribution is
broader than the raw one, with the fractional error of the fitted
$\sigma$ dependent roughly on the ``fractional error" (with respect
to $\sigma$) of the first moment, i.e. increasing as the distribution
becomes sharper.

Since the Wigner distribution is not symmetric
about its peak, the corresponding error in fitting an off-center
CGWD by a properly centered CGWD should not be depend purely quadratically
on $\Delta \mu_1$.  Indeed, we find over the range 1 $<\varrho <$ 8
that an excellent approximation is
\begin{equation}
\Delta\varrho/\varrho
\approx (0.3\varrho-3.0)\Delta\mu_1+(-2.0\varrho+0.4)(\Delta\mu_1) ^2 \ .
\end{equation}

Analogous to the previous result for $\sigma$, the fractional error
of $\varrho$ has strong quadratic tendencies, with the magnitude of the
curvature increasing with increasing $\varrho$.  The linear term
complicates
behavior, causing $\varrho$ to increase for small shifts of the curve
to the right.  Evidently for some $\varrho$-dependent offset, the best
fit will coincidentally give the true value of $\varrho$.

\section{Wigner Distribution as a 2-Parameter Fit}
\label{sec-wigner2param}

In fitting experimental TWDs, it
becomes apparent that in many cases --- particularly when the data are
relatively poor --- the CGWDs giving the best fits have
first moments different from the first moments
of the data.  GE noted that
the peak of TWDs can be well fitted by treating
a Gaussian as a 3-parameter fitting function, with the peak
position and the prefactor allowed to vary along with the standard
deviation.  (Presumably the prefactor differs from its expected value,
set by normalizing the Gaussian, because of the existence of a small
``hump'' sometimes observed at large values of $s$ [see below].)
In contrast, it is not clear how such arbitrary modifications
could be made to the CGWD, nor is it clear what physical information
could be extracted from a CGWD with arbitrary modifications.

{}From a basic perspective, though, it might be desirable to
determine the scaling length (the ``effective mean,'' which
equals the first moment for ideal CGWDs) and the
variance in a single fitting procedure rather than to find this length first
from the first moment or otherwise.  For the following discussion, we
denote by $\bar{\ell}$ the effective mean determined
as one parameter of a two-parameter least-squares fit of the data to
a CGWD, the
other parameter being the exponent $\varrho$.
This refined scaling implies that the argument of $P_{\varrho}$ should be
$\ell/\bar{\ell}$. It is convenient to introduce a new adjustable parameter
$\alpha$ which gives the ratio of $\bar{\ell}$ to the actual mean step
separation
$\langle \ell \rangle$.  Since $s$
--- still defined as $\ell/\langle \ell \rangle$ ---
is the natural variable to use in describing data,
our refined scaling translates into replacing
$s$ by $s/\alpha$ in the argument of the distribution.  If the integration
variable $s$ were also replaced by $s/\alpha$, then the refined scaling would
amount to a redefinition of a dummy variable, and normalization
would still be realized.  Since the independent variable is kept as
$s$, the extra factor is associated with $P(s)$ instead:

\begin{equation}
\check{P}_{\varrho,\alpha}(s) \equiv \frac{P_{\varrho}(s/\alpha)}{\alpha}
\label{e:Psa}
\end{equation}

\noindent In other words, the first moment of the distribution, $\mu_1 \equiv
\langle \ell \rangle$ occurs at 1/$\alpha$ times the optimal characteristic
terrace width $\bar{\ell}$.

We used {\it Mathematica}$^{\bigcirc\hspace{-0.6em}{\rm R}}$
regression routines to fit the experimental
data by minimizing the value of  $\chi^{2}$ as a function of the
adjustable parameters. Since the values of
$s$ are quantized (cf.\ Sec.~\ref{sec-discrete}), there was assumed
to be
no error in these values.  For simplicity, all data points were weighted
equally.

\subsection{Copper: Moderately Strong Repulsions}

Our findings for vicinal Cu surfaces are presented in Table
1, which is similar to Table 2 of GE, but contains many cases
of ``poor'' data omitted in GE.  In order to facilitate discussion,
TWDs were divided by GE into
three groups based on a visual assessment of their quality:
\begin{itemize}
  \item A ``good'' TWD changes height essentially monotonically below
	the peak and again above it; there are no dips, humps, or
	double peaks, and there is minimal scatter in the data points.
	``Good'' data are indicated by a ``$+$'' in Table 1.
  \item An ``OK'' TWD has more scatter, with small dips and peaks
	introduced by variations (within the limits of the
	general margin of error) of single data points.
	``OK'' data are indicated by a ``$0$'' in Table 1.
  \item A ``poor'' TWD has a double-peak or hump at large $s$;
	correspondingly, the position of the (main) peak occurs
	noticeably below $s$ = 1, even when the peak is fairly
	narrow and the skewness minimal.
	The judgment that this data is ``poor'' is based both on
	the intuition of the {\em experimenter} and on the following
	argument:
        A second peak at
	large $s$ would be characteristic of the onset of faceting;
	however, ``poor'' data tends to occur at high temperatures,
	whereas faceting should be more important at low temperatures.
	``Poor'' data are indicated by a ``$-$'' in Table 1.
\end{itemize}

As expected, the Gaussian distribution
yields a reasonable, but not exceptional, fit to the data; it
worked especially well on surfaces with low temperatures, so
relatively large $\tilde{A}$. As an example of good data---exemplified by
the
vicinal (1~1~13) surface at 300K, depicted in Fig.~1---the
[three-parameter]
Gaussian yields a $\chi^2$ value of about 0.0072. The single-parameter
CGWD fit gives a slightly worse fit to the data, having a $\chi^2$  of
0.0078. For the two-parameter Wigner fit (Eq.\ (\ref{e:Psa})), the
$\chi^2$
value improves by better than a factor of  two, to 0.0037, with a value of
$\varrho$ increased slightly (from 6.4  to 6.5, leading
to a value of $\sigma$ closer to that from the Gaussian fit.
In this case, the optimal fit using
Eq.\ (\ref{e:Psa}) is obtained by scaling the terrace widths with a value
that is 96.5\%
of that given by the first moment of the distribution.
In other  words, the first moment of the TWD is 3.6\%
greater than the
value of the mean spacing associated with the best fit of the
distribution.

In Fig.~2, we display results for this same vicinal Cu surface at 378K as an
example of poor data, with a large shift in the  effective mean.  In this
case, having extra degrees of freedom in the fit makes a sizable
difference. For the three-parameter Gaussian fit, the $\chi^2$ is 0.035;
$\chi^2$ increases to 0.042 for the single-parameter Wigner fit and to
about half that value, 0.025, for the two-parameter fit, all these values
being half an order of magnitude larger than in the previous, good case.
The
value of $\varrho$ increases noticeably --- from 2.5 to 3.0 --- when the
refined scaling is allowed (and rises to 4.3 for the shifted-mean method).  The
refined scaling factor for terrace widths is 0.867, meaning that the
explicit average
$\langle \ell \rangle$ of the TWD
is 15.3\%
greater than the  value of the mean spacing associated with the
best fit of the  distribution.  Characteristic of this sort of data is the
hump on the high-$s$ side of the peak, which distorts the single-parameter
CGWD fit so that it poorly reproduces the peak region.

We emphasize the following general trends in Table~1:  In almost all instances,
the value of $\bar{\ell}$ derived from the two-parameter fit to  a CGWD is
smaller
than $\mu_1= \langle \ell \rangle$ given by the first moment (the average)
of the
TWD; likewise, the directly measured values of $\sigma$ are almost always
larger
than the values obtained by any of the three fitted curves.
(Cf.\ Sec.~\ref{sec-sample}.)
The value of
$\varrho$ is higher for the scaled fit than for the  single-parameter CGWD fit,
and the associated value of
$\sigma$ typically closer to that deduced from the Gaussian fit.  For
``good'' data, the
change of value of $\mu$ is of order a few percent, and the change in $\varrho$
and $\sigma$ is negligible.  For ``poor" data, the refined scaling factor
is at least
twice as large and the two-parameter-fit curve is narrower than the
single-parameter-fit curve.
The tails or humps in
the experimental TWDs seem to be responsible  for the systematic discrepancies
in the fits, especially the smaller mean and smaller variance of the fits
relative to the direct measurements.

\subsection{Platinum: Weak Repulsions}

We have also considered recently reported data for vicinal Pt(110)
at room temperature\cite{SBV}.  In this system the terraces are
($1 \! \times \! 2$)
reconstructed, and the steps correspond to 3-unit segments (as
would be found in a ($1\! \times \! 3$) reconstruction).  The authors in that
paper conclude that the interaction between their steps is small, but
are unable to proceed to a quantitative assessment using preexisting
methods: Gaussian methods are utterly inappropriate for this
regime of small interactions.

In Fig.~3, we show single- and two-parameter
Wigner fits of the data.  For the former,
$\varrho$ = 2.06 ($\tilde{A}$ = 0.0309), with a
$\chi^2$ of 0.008.  With the latter, the optimal $\bar{\ell}$
for determining $s$ is 91\%
of $\langle \ell \rangle$ predicted by the
average of the data (viz.\ $\alpha$ = 0.91);
$\varrho$ rises to 2.24 ($\tilde{A}$ = 0.134)
and the quality of the fit improves to $\chi^2$ = 0.003.

Thus, the high-$s$ bulge does not seem to be peculiar to the vicinal
Cu systems of GE.
We do not understand the physical origin of the systematic need
for refined scaling of experimental data.  We see no comparable effect
in our companion Monte Carlo simulations, reported elsewhere\cite{CRE}.

\section{Effects of discreteness on continuum models of a TWD}
\label{sec-discrete}

Due to the crystalline nature of the surface, the TWD is a discrete
rather than  a continuous function: the TWD should have a sizable
number of counts only at values of $\ell$ that are
$a_{\perp}$ times the sum of an integer and a constant
fractional offset determined by the terrace and the orientation of
the steps.  (E.g., this offset is 1/2 for close-packed steps on
\{1 0 0\} surfaces of fcc crystals.)  For simplicity we neglect this
offset in this paper, setting it to zero (as on \{1 0 0\} surfaces of
sc crystals). Thus,
$s$ can only take on the values $s_L \equiv L
a_{\perp}/\langle\ell\rangle \equiv L/\langle L \rangle$, $L$ being a
positive integer. It is very tempting simply to apply formulae derived
for  the continuous TWD given by Eq.~(\ref{e:Wigner}). In this section
we discuss the potential difficulties posed by the discrete nature of
the TWD. Inspired by the scaling of discrete TWDs \cite{joos}, we construct
a {\it discrete} generalized Wigner distribution (DGWD) TWD given by
\begin{equation}
  \label{e:discreteWigner}
   \breve{P}_\varrho(s) =  \breve{a}_{\varrho} s^\varrho \exp
\left(-b_\varrho s^2\right) \sum_L \delta(s - s_L),
\end{equation}
where $\breve{a}_{\varrho}\! \approx \! a_{\varrho}/\langle L
\rangle$ is determined by the requirement of normalization.

Although $b_\varrho$ was defined so as to make the mean of
the CGWD unity, there is no guarantee that the same parameter
will make the mean of the DGWD unity; likewise,
the two functions may have different variances.
We chose values of $\langle L \rangle$ and $\varrho$ to specify a DGWD and
then numerically performed two-parameter fits using CGWD formulae
[Eqs.~(\ref{e:Wigner})--(\ref{e:abr}), (\ref{e:Psa})] to produce estimates
of $\varrho_c$.  Anticipating greater interest in behavior as a function
of
$\tilde{A}$ than of $\varrho$, we converted our results for $\varrho_c$ to
$\tilde{A}_c$ using Eq.~(\ref{e:abr}).

Fig.~\ref{discrete_rho} shows the difference of the
fitted  value $\tilde{A}_c$ and the ``parent" value $\tilde{A}$ as a function
of this $\tilde{A}$ for
several mean widths $\langle L \rangle$.   As may be expected, as the TWD
becomes narrower (i.e.\ for sufficiently large $\tilde{A}$ or
$\varrho$), $\tilde{A}_c$ becomes an decidedly unreliable estimate for
$\tilde{A}$; based on examination of the cases $\langle \ell
\rangle/a_{\perp}$ = 2--6, this breakdown appears to occur for $\varrho$
near
$\langle L \rangle^2$.  This threshold corresponds to $s_{L+1}-s_L \!
\equiv \! a_{\perp}/\langle \ell \rangle \! \equiv \! \langle L
\rangle^{-1} \! = \! \sigma$. Thus, for $\langle L \rangle \! < \! 4$,
this breakdown occurs in the region of physical  interest (cf.\ dashed
curves in Fig.~\ref{discrete_rho}). On the other hand, for $\langle L
\rangle \! \geq \! 4$,
$\tilde{A}_c$ provides a reasonable estimate of
$\tilde{A}$ over the range of physically reasonable
dimensionless repulsions,  where the effects of discreteness are most
pronounced for small values of $\langle L \rangle$ of $\tilde{A}$.  Note
also that
in
each case there is a substantial peak in $|\tilde{A}_c - \tilde{A}|$ for
small $\tilde{A}$. Fig.~\ref{discrete_l0} shows the reduction in the error
in $\tilde{A}_c$ as $\langle L \rangle$ increases, at fixed values of
$\tilde{A}$.

In summary, we have raised a flag of caution when analyzing the
fluctuations of
highly misoriented vicinal surfaces in a conventional framework.  The case
of $\langle L \rangle = 3$ corresponds to (1,1,7) for close-packed steps on
surfaces
vicinal to
\{1 0 0\} planes of fcc crystals. Thus, one should view with some
suspicion
the unusually large values of $\varrho$ and
$\tilde{A}$ reported for the single temperature at which this vicinal Cu
surface was measured.  For \{1 1 1\} surfaces, the corresponding Miller
indices are (5 3 3) for A steps (\{1 0 0\} microfacets) and (2 2 1) for B
steps (\{1 1 1\} microfacets)\cite{EE}.

We also emphasize that this behavior is not a vagary of Wigner
distributions.  Misorientation causes similar problems when the mean and
variance
of discretized Gaussian TWDs are analyzed as though they were continuous
Gaussian
functions.    For more convenient comparison with the above Wigner
distribution,
we used Eq.~(\ref{e:sr}) to relate the variances and values of
$\varrho$.  We found that estimates of $\varrho$ based on the
variance of the discretized Gaussians approached the undiscretized
value monotonically, rather than oscillating as in the case of
the Wigner distribution, and that the approach to the undiscretized
value of $\varrho$ is actually somewhat {\em slower} in the Gaussian
case than in the Wigner case.  The Gaussian case also showed a
breakdown at large values of $\varrho$ (small $\sigma^2$) similar to
the Wigner case.

\section{Statistical uncertainties due to finite sampling size}
\label{sec-sample}

By truncating Eq.~(\ref{eq:rho_of_sig2}) at the second term,
we can create an estimator $\hat{\varrho}$ for $\varrho$:
\begin{equation}
  \hat{\varrho} =
     \frac{1}{2}\left(\widehat{\sigma^2}\right)^{-1} - \frac{3}{4}\,
,
\end{equation}
where $\widehat{\sigma^2}$ is a random variable that is an estimator of
$\sigma^2$.
For small $\sigma^2$, though, the Wigner distribution approaches
a more familiar Gaussian distribution, as discussed in
Sec.~\ref{sec-normmean}.
For a Gaussian distribution, the sampling errors from a sample
of size $N_{\rm samp}$ for
$\widehat{\sigma^2}$ are given by\cite{VarSampErr}
\begin{equation}
  \label{e:samperr}
   \mbox{var}\left(\widehat{\sigma^2} \right)
     = \frac{2\sigma^4}{(N_{\rm samp}-1)} \, .
\end{equation}
Accordingly, the standard deviation of the estimated values of
$\varrho$ can be seen to increase with increasing values of $\varrho$:
\begin{equation}
  \sqrt{\mbox{var}(\hat{\varrho})} = \sqrt{\frac{2}{N_{\rm samp}-1}}
        \left( \varrho + \frac{3}{4} \right) \, .
\label{e:rtvarrho}
\end{equation}

In this section we explore the effects of statistical fluctuations
on the estimated value of $\varrho$ by performing some well-defined numerical
experiments.  The results are thus applicable to ``ideal" data.  In fact there
apparently are systematic effects, noted earlier, in real data that limit the
applicability of some deductions.

Specifically, we begin with the following simple
procedure: First, we independently choose $N_{\rm samp}$ values of $s$
using
the same known DGWD as the probability density function for each
selection.
Second, we fit this artificial TWD using the two-parameter Wigner
distribution $\check{P}_{\varrho,\alpha}(s)$ to determine
$\varrho$, taking each point to be weighted equally in accordance
with standard practice\cite{GE}.
Third, we repeat this process a large number of times
and measure the standard deviation of the fitted values of
$\varrho$ as well as any systematic bias in the fitted estimates.

Fig.~\ref{fig:sample_dRho} shows the result of this
procedure, with one million independent TWDs produced for each
value of $\varrho$ and each TWD consisting of 801 independent
values of $s$ drawn according to a DGWD.
Clearly  the linear relationship between the
$\sqrt{\mbox{var}(\hat{\varrho})}$ is maintained, but
the slope is somewhat larger than predicted by
Eq.~(\ref{e:rtvarrho}).

Another way of estimating $\varrho$ is to measure directly the
mean and variance of the TWD and to insert them into
Eq.~(\ref{eq:rho_of_sig2}).  Repeating our procedure of
creating artificial TWDs, we accordingly estimate $\varrho$ using
Eq.~(\ref{eq:rho_of_sig2}), again
analyzing  the variance of the estimates as above.  As
seen in Fig.~\ref{fig:sample_dRho}, the resulting estimates of
$\varrho$ have variances given almost exactly by
Eq.~(\ref{e:rtvarrho}) and noticeably smaller (though not by a large
factor) than the variances given  by the traditional,
uniformly-weighted nonlinear least-squares fits.  This finding means
that not only is it possible to use simple analytic functions to
find $\varrho$ and $\tilde{A}$ instead of using  two-parameter
nonlinear least-squares fits, but also that doing so is
statistically better!

This result appears to be  contrary to the belief that
performing a least-squares fit to an appropriate smooth function
is desirable to minimize the effects of statistical fluctuations.
It seems likely, though, that the real problem lies in the weighting
of the data in the fit.  It has been suggested that greater weight
should be given to the points near the peak of the TWD, so we once
again repeat our procedure, this time making a least-squares fit in
which each point is weighted proportionally to the measured value of $P$.
As Fig.~\ref{fig:sample_dRho} shows, the standard deviation of
$\hat{\varrho}$ again varies linearly with $\varrho$, but with a slope
that is slightly {\it higher} than that of the uniformly-weighted
case.  In retrospect, this result should not be surprising, since
each point on the TWD represents the result of $N_{\rm samp}$
binomial experiments ({\em i.e.,} Bernoulli trials: either the  measurement
of step separation gives {\em this} distance $s_L$ or {\em some other}
distance).  Elementary statistics\cite{MendenhallMatStat} shows that
the statistical error of binomial experiments is smallest when
the probability of success is nearly zero or nearly one --- in
our case, for points on the TWD with
$P(s)\!\approx\!0$.  However, devising a naive weighting by the
reciprocal of the  variance of each point on the TWD
is problematic when points for which the measured value of
$P(x)$ is equal to zero; these points would receive {\em infinite}
weight, yielding nonsense results.
Even if one circumvents this problem, there is still the problem that the
points are not uncorrelated, which is a requirement of the
least-squares procedure\cite{StatisticsForExperimenters};
the normalization condition imposes
a (weak) correlation between points.
We can avoid these problems by simply
using the mean and variance of the TWD in Eq.~(\ref{eq:rho_of_sig2})
to find
$\varrho$ or Eq.~(\ref{e:as}) to find $\tilde{A}$.

Motivated by these numerical experiments, we computed directly (from the
histogram data, after normalization and adjustment to unit mean) the standard
deviation $\sigma_{\rm dir}$ (cf. Table~1).  This result is systematically
higher than the
$\sigma$'s obtained from the various fitting techniques.  The difference is
modest for good data but pronounced for poor data.  Thus, as mentioned in
Sec.~\ref{sec-wigner2param}, it appears to come from the curious high-$s$
undulations that plague poor data.  The fitting techniques, being less
sensitive
to these points, give values of $\sigma$ that are less distorted by them.

Finally, we note that single STM images do not allow for a large
number of independent values of $s$.
The number of independent measurements is generally
much smaller than the total number of measurements, due to
correlations between the measurements.
Although a precise determination of the effects of correlations
on fitted parameters would be rather involved, a working
estimate of the number of ``independent'' measurements
--- from which uncertainties can be estimated ---
can be made in the following way.
First, one obtains the terrace width $\ell_n(y)$ between
steps $n$ and $n\! + \! 1$ for each position $y$
along the steps.  Then the correlation
function\cite{StatisticsForExperimenters}
\begin{eqnarray}
   C_n(y) & \equiv &
	\frac{\langle \ell_0(0) \ell_n(y) \rangle - \langle \ell \rangle^2}
             {\langle \ell^2 \rangle - \langle \ell \rangle^2} \nonumber \\
 & = &  \frac{1}{\langle \ell^2 \rangle - \langle \ell \rangle^2}
         \left[\frac{\sum_{n^\prime}^{N-n}\sum_{y^\prime=1}^{L_y-y}
                    \ell_{n^\prime}(y^\prime) \ell_{n^\prime+n}(y^\prime+y)}
                   {(N-n)(L_y-y)}
          - \langle \ell \rangle^2\right]
\end{eqnarray}
is calculated, where $N$ is the number of terraces in the image
({\em i.e.,} $N+1$ is the number of steps).
The correlation function along the steps decays exponentially
as $C_0(y) \sim \exp(-y/\xi_y)$,
where $\xi_y$ is the intrastep correlation length.\footnote{
As discussed in Ref.~\cite{BEW}, from the
Gruber-Mullins\cite{Gruber67} perspective
$\xi_y = \frac{8 \langle \ell \rangle^2 \tilde{\beta}}
              {3 \pi^2 k_{\rm B} T}$
for $A \! = \! 0$ and
$\xi_y \approx \frac{2 \langle \ell^2 \rangle \tilde{\beta}}
                    {k_{\rm B} T}$
for $A \! \gg \! 0$.
}
$\xi_y$, is given by Eq.~(18) of Ref.~\cite{BEW}, but the
safest procedure is simply to measure it.
The correlation function between steps, on the other hand,
is more complicated; $C_1(0)$ is negative\cite{MehtaRanMat},
but the trend is for the absolute value of
$C_n(0)$ to decrease rapidly with increasing $n$.
We define $y_c$ to be the smallest value of $y$ for which
\begin{equation}
  |C_0(y)| \leq c \quad \forall \, y \geq y_c\, ,
\end{equation}
and likewise $n_c$ to be the smallest value of $n$ for which
\begin{equation}
  |C_n(0)| \leq c \quad \forall \, n \geq n_c\, ,
\end{equation}
where $c$ is a small cutoff (we recommend $c \! = \! 0.1$).
The number of ``independent'' terrace widths will be approximately
$(L_y/y_c)(N/n_c)$.

As an example, we performed a Monte Carlo simulation of
a system with $A \! = \! 0$.
For simplicity, we chose $k_B T$ to be equal to the energy for
producing a kink, and we chose the mean distance between steps to
be ten lattice units.   We measured $\xi_y \approx 15$
(consistent with theoretical predictions, see precedingy footnote) and
$y_c \approx 40$, and we
observed that $|C_n(0)| \leq 0.1$ for all $n \geq 3$.  Suppose this
had been an STM image representing a square region of the crystal
200 lattice units on a side; then there would be approximately
20 terraces in the image, and  $(200/40)(20/2) = 50$ independent
widths --- much smaller than the 4000 independent measurements that
one might naively suspect.  As a result, we see that the uncertainty
in statistically derived quantities such as the measured value of $A$
are an order of magnitude larger than the naive estimate.
Lowering the temperature relative to the kink creation energy would
have the effect of further reducing the number of independent
measurements and thus increasing the uncertainty in the measured value
of $A$.

With such small samples, the measured TWD can differ distinctly from
the DGWD due to statistical fluctuations alone.
In order to demonstrate this idea, we produced 20 TWDs, each
consisting of 400 independent values of $s=s_L$ sampled from a DGWD with
$\varrho\! = \! 5$ and $\langle L \rangle \! = \! 6$,
and fitted each TWD with a DGWD.  (In this model,
$s_L = L/\langle L \rangle$; there is no offset between
successive terraces.)
Fig.~\ref{samplepics} shows the TWDs with the lowest and the highest values of
$\chi^2$.  Curiously, in this particular case the TWD with
the largest value of $\chi^2$ happens to
produce better estimates of both $\varrho$ and $\langle L \rangle$
than does the TWD with the smallest value of $\chi^2$.  In no case,
however, do we see the shoulders or second peaks
in the TWD at
large values of $s$ as occur {\em systematically} in the ``poor" data of
Fig.~\ref{Cupoor} here or Fig.~5b in Ref.~\onlinecite{GE}.
Since the ``poor'' data were based on several dozen independent
STM measurements, they should be statistically comparable to the
data of Fig.~\ref{samplepics}, but the systematic deviation
indicates that the ``poor" data cannot be entirely understood
within the framework of
a generalized Wigner distribution.

\section{Conclusions}
\label{sec-conclusions}
In this paper we have performed several numerical experiments and analyses to
understand better the TWDs derived from physical data from
vicinal surfaces.  We have quantitatively studied how the Wigner
distribution approaches a
Gaussian for large dimensionless
interactions, and shown that for most systems of physical interest
the standard deviation of the terrace width can be estimated from
either distribution with little difference.

The mean step width can be estimated in a variety of ways for
both Wigner and Gaussian distributions;  two reasonable
but inequivalent choices
are directly averaging the step width and fitting the TWD to
the desired distribution using a nonlinear least-squares routine.
In Sec.~\ref{sec-normmean} we discuss the effects of using an
estimated mean that differs from the least-squares estimate.
On the other hand, the adjustment of
the normalization of the curve to obtain a "better" fit is unjustified;
even if the visual agreement appears to improve, no
results from theory --- such as interaction strengths ---
can be meaningfully extracted from TWDs with the wrong normalization.

We have proposed a two-parameter extension of the
generalized Wigner surmise, which really is just a consistent
fitting of both $\varrho$ and the mean terrace width within a
single two-parameter least-squares fit.
This added flexibility allows one to deal more fruitfully with
poorer-than-desirable experimental data, while not changing good data (or
the data emerging from various numerical simulations).  Thus, this fitting
function can applied generally.
On the other hand, for ``good'' data, we have shown that a simple
series expansion based on the directly measured mean and standard
deviation of the terrace widths has better statistical properties.
The least-squares fit is more robust, but this is a property that
is useful {\em only} when the Wigner distribution does not capture
all of the important physics, such as the role of defects or
more complicated step-step interactions; in such cases, the
estimated values of the step-step interaction from the
Wigner --- or the Gaussian --- distribution
must be treated with great caution.

Finally, we emphasize the importance of using many STM measurements to
insure good statistics and the desirability of calculating the correlations
between terrace widths within individual STM images.  As we saw in
section~\ref{sec-sample}, the typical STM image suitable for measuring
terrace widths will contain no more than about 50 {\em independent}
terrace width measurements, almost two orders of magnitude less than the
4000 total terrace width measurements.

\section*{Acknowledgement}

Work at Maryland was supported by NSF-MRSEC
Grant No.\ DMR-9632521 and benefited from interactions with
E.~D.\ Williams.  SDC, in particular, participated in an
NSF-MRSEC-sponsored REU program.  We are grateful to K.~Swamy
and E.~Bertel for sending their data for vicinal Pt(110). TLE
also acknowledges the support of a Humboldt U.S.\ Senior Scientist
Award and  the hospitality of the IGV at FZ-J\"{u}lich.

\clearpage
\section*{Appendix A: Derivation of Expansions}
\setcounter{equation}{0}
\renewcommand{\theequation}{A \arabic{equation}}
In this appendix we derive a useful series expansion Eqs.\ (\ref{e:br}) and
(\ref{e:method2}) for  the coefficient $b_\varrho$ in the quadratic
exponential of the generalized Wigner distribution.
For convenience, we define the variable
\begin{equation}
  r \equiv \frac{\varrho +1}{2} \,.
\end{equation}
Then we use Stirling's asymptotic series \cite{AS},
$\Gamma(r) = \pi^{1/2}\exp(-r) r^{r -1/2}F(r)$:
\begin{eqnarray}
    b_\varrho & \equiv &\left[\frac{\Gamma \left(r+\frac{1}{2}\right)}
                           {\Gamma \left(r            \right)}\right]^2
                \nonumber \\
& = & \left[
            \frac{e^{-(r + 1/2)} (r+ 1/2)^{r}F(r+1/2)}
                 {e^{-r} (r     )^{r    -1/2}F(r    )}
            \right]^2 \, ,
\end{eqnarray}
where
\begin{equation}
  F(r) = 1 + \frac{1}{12 r} + \frac{1}{288 r^2}
         - \frac{139}{51840 r^3} - \frac{571}{2488320 r^4}
         + {\cal O}(r^{-5}) \, .
\end{equation}

We concentrate initially on the first part of the fraction:
\begin{equation}
\left[
  \frac{e^{-(r + 1/2)} (r+ 1/2)^{r}}
       {e^{-r      } r^{r    -1/2}}
\right]^2
  = e^{-1} r  \left(1 + \frac{1}{2r}\right)^{2r} \, .
\end{equation}
But
\begin{eqnarray}
  \left(1 + \frac{1}{2r}\right)^{2r} & = &
   \exp\left\{ \ln \left[
         \left(1 + \frac{1}{2r}\right)^{2r}
      \right] \right\} \nonumber \\
 & = &
   \exp\left\{
         \left[  1
               - \frac{1}{2} (2r)^{-1}
               + \frac{1}{3} (2r)^{-2}
               - \frac{1}{4} (2r)^{-3}
               + \frac{1}{5} (2r)^{-4}
                  + {\cal O}[(2r)^{-5}]
       \right]
      \right\} \nonumber \\
 & = & e^{+1}\left[         1
      - \frac{1}{2}         (2r)^{-1}
      + \frac{11}{24}       (2r)^{-2}
      - \frac{7}{16}        (2r)^{-3}
      + \frac{2477}{5760}   (2r)^{-4}
                 + {\cal O}[(2r)^{-5}]
       \right] \, .
\end{eqnarray}
It is also straightforward to show that
\begin{equation}
 \left[\frac{F(r+1/2)}{F(r)}\right]^2
   =                    1
     - \frac{1}{3}     (2r)^{-2}
     + \frac{1}{3}     (2r)^{-3}
     - \frac{13}{90}   (2r)^{-4}
            + {\cal O}[(2r)^{-5}] \, .
\end{equation}
By combining all of these we find
\begin{equation}
  b_\varrho  = r \left( 1
                   - \frac{1}{4}
                   + \frac{1}{32}  r^{-1}
                   + \frac{1}{128}   r^{-2}
                   - \frac{13}{6144} r^{-3}
                         + {\cal O}[r^{-4}]
             \right) \, .
\end{equation}
{}From this formula for $b_\varrho$, we can use Eq.~(\ref{e:sra}) to write
$\sigma^2$ as a power series in $\varrho^{-1}$:
\begin{eqnarray}
  \sigma^2  & = & \frac{1}{2}   (\varrho + 1)^{-1}
                + \frac{1}{8}   (\varrho + 1)^{-2}
                - \frac{1}{16}  (\varrho + 1)^{-3}
                - \frac{17}{384}(\varrho + 1)^{-4}
                     + {\cal O}[(\varrho + 1)^{-5}] \nonumber \\
            & = & \frac{1}{2}   \varrho^{-1}
                - \frac{3}{8}   \varrho^{-2}
                + \frac{3}{16}  \varrho^{-3}
                + \frac{7}{384} \varrho^{-4}
                     + {\cal O}[\varrho^{-5}]
             \, .
\end{eqnarray}
Using reversion of series, we then find
\begin{equation}
  \varrho^{-1} =  2  \sigma^2
             + 3 (\sigma^2)^2
             + 6 (\sigma^2)^3
             + \frac{32}{3} (\sigma^2)^4
                     + {\cal O}[(\sigma^2)^{5}] \, ,
\end{equation}
from which we get
\begin{eqnarray}
  \varrho & = &\frac{1}{2}\left(\sigma^2\right)^{-1}
           \left\{ 1
             + \frac{3}{2} (\sigma^2)
             + 3 (\sigma^2)^2
             + \frac{16}{3} (\sigma^2)^3
                     + {\cal O}[(\sigma^2)^{4}]  \right\}^{-1} \nonumber
\\
   & = & \frac{1}{2}\left(\sigma^2\right)^{-1}
           \left\{ 1
             - \frac{3}{2} (\sigma^2)
             - \frac{3}{4} (\sigma^2)^2
             + \frac{7}{24}(\sigma^2)^3
                     + {\cal O}[(\sigma^2)^{4}]  \right\} \, .
\end{eqnarray}
Finally, with Eq.~(\ref{e:ar}) we can use these results to find the
dimensionless interaction constant $\tilde{A}_W$ in terms of $\sigma^2$:
\begin{equation}
  \label{e:expand_Atilde}
  \tilde{A}_W
    =  \frac{ 1}{16} (\sigma^2)^{-2}
          - \frac{ 7}{16} (\sigma^2)^{-1}
          + \frac{27}{64}
          + \frac{35}{96} (\sigma^2)
          + {\cal O}[(\sigma^2)^{2}]  \, .
\end{equation}
For $\tilde{A} \! \geq \! 0.0525$, the relative error in
Eq.~(\ref{e:expand_Atilde}) is less than
1\% (less than 0.1\% for
$\tilde{A} \! \geq \! 0.15$).
The absolute error is less than 1\% for
$\tilde{A} \! \geq \! -1/4$.

\newpage
\section*{Appendix B: Effect of displacement of a Gaussian fitting
function
in fits of a Gaussian}
\setcounter{equation}{0}
\renewcommand{\theequation}{B \arabic{equation}}

Suppose we have a Gaussian distribution with mean $\mu$ and variance
$\sigma^2$; we attempt to fit this Gaussian with a second Gaussian
with a mean $\mu + \Delta \mu$ and a variance $(1-\zeta)^2\sigma^2$,
where $\Delta \mu$ is fixed and $\zeta$ is unknown.   Explicitly, we
write
$\chi^2$ as a function of $\zeta$ and $\Delta \mu$:
\begin{eqnarray}
 \chi^2 & = & \int_{-\infty}^{\infty}{\rm d}x \Biggl\{
              \left[\frac{1}{\sigma (1-\zeta) \sqrt{2\pi}}
\exp\left(-\frac{(x-\mu)^2}{2\sigma^2(1-\zeta)^2}\right)\right]
                \nonumber \\
       & & \mbox{} -
              \left[\frac{1}{\sigma \sqrt{2\pi}}
                    \exp\left(-\frac{[x-(\mu+\Delta \mu)]^2}
                       {2\sigma^2}\right)\right]\Biggr\}^2
             \nonumber \\
       & = & \frac{1}{2\sigma(1-\zeta)\sqrt{\pi}}
            + \frac{1}{2\sigma\sqrt{\pi}} \nonumber \\
 & & \mbox{} + \frac{1}{\sigma\sqrt{2\pi[1+(1-\zeta)^2]}}
      \exp\left[-\frac{(\Delta \mu)^2}
                {2\sigma^2 \{1+(1-\zeta)^{2}\} }\right] \nonumber \\
 & = & \frac{1}{\sigma\sqrt{\pi}} \Biggl(
       \left[\left(\frac{\Delta \mu}{2\sigma}\right)^2
              + {\cal O}\left\{\left(\frac{\Delta
\mu}{2\sigma}\right)^4\right\}
          \right] \nonumber \\
 & & \mbox{} +
       \left[\frac{3}{2} \left(\frac{\Delta \mu}{2\sigma}\right)^2
             + {\cal O}\left\{\left(\frac{\Delta
\mu}{2\sigma}\right)^4\right\}
          \right]\zeta
\label{e:X1} \\
 & & \mbox{} + \frac{3}{8}
       \left[1 +
          3 \left(\frac{\Delta \mu}{2\sigma}\right)^2
          + {\cal O}\left\{\left(\frac{\Delta
\mu}{2\sigma}\right)^4\right\}
          \right]\zeta^2 + {\cal O}(\zeta^3)\Biggr) \, . \nonumber
\end{eqnarray}
Since the optimum fit is found by minimizing $\chi^2$, we set
${\rm d}\chi^2/{\rm d}\zeta \! = \! 0$ in Eq.\ (\ref{e:X1}) and remove
the overall prefactor $1/(\sigma\sqrt{\pi})$ to get:
\begin{eqnarray}
0 & = & \left[\frac{3}{2} \left(\frac{\Delta \mu}{2\sigma}\right)^2
             + {\cal O}\left\{\left(\frac{\Delta
\mu}{2\sigma}\right)^4\right\}
          \right]  \\
 & & \mbox{} + \frac{3}{4}
       \left[1 +
          3 \left(\frac{\Delta \mu}{2\sigma}\right)^2
          + {\cal O}\left\{\left(\frac{\Delta
\mu}{2\sigma}\right)^4\right\}
          \right]\zeta + {\cal O}(\zeta^2) \, . \nonumber
\end{eqnarray}
Solving for $\zeta$, we find
\begin{equation}
  \zeta = 2 \left(\frac{\Delta \mu}{2\sigma}\right)^2
     + {\cal O}\left\{\left(\frac{\Delta \mu}{2\sigma}\right)^4\right\} \,
.
\label{e:zfin}
\end{equation}
Since $\Delta \sigma^2/\sigma^2 = 2\Delta \sigma/\sigma$,
Eq.~(\ref{e:zfin}) leads to Eq.~(\ref{e:Ds}).

\clearpage
\begin{figure}[t]
\caption[shrt]{
The vicinal surface Cu(1 1 13) at 300K is an example of good data.  The
points show the  normalized data from the STM image. The short-dashed curve
shows the conventional (three-parameter) Gaussian fit to the data;
the fitted standard deviation is $\sigma_G = 0.25 \pm 0.01$.
 The long-dashed curve shows a fit to a generalized Wigner distribution
with the exponent $\varrho$ as the single adjustable parameter. The
best-fit result is $\varrho = 6.4 \pm 0.5$, leading via
Eq.~(\ref{e:sra}) to the estimate $\sigma_W = 0.26 \pm 0.01$. The
terrace widths are scaled by the mean spacing determined from the
average of the data.  In the solid curve, the Wigner distribution is
treated as a two-parameter function.  We now find $\varrho = 6.5 \pm
0.3$, leading again to $\sigma_W = 0.26 \pm 0.01$.
}
\label{Cugood}
\end{figure}

\begin{figure}[t]
\caption[shrt]{
The vicinal surface Cu(1 1 13) at the higher temperature 378K is an
example of
poor data.  As in Fig.~\ref{Cugood} the points show the data, the
short-dashed curve a conventional Gaussian fit, the long-dashed curve a
single-parameter Wigner fit, and the solid curve a two-parameter Wigner
fit. For the Gaussian fit, we get $\sigma_G = 0.30 \pm 0.04$, a
broader distribution than in Fig.~1, as expected for the higher T. In
contrast to Fig.~\ref{Cugood}, there is a considerable difference
between the two Wigner fits, with the two-parameter version
providing a much better accounting due to its ability to adjust to
accommodate the points near the peak. For the one-parameter fit, we
find  $\varrho = 2.5 \pm
0.7$, leading to $\sigma_W = 0.39 \pm 0.03$, while for the
two-parameter fit, we get  $\varrho = 3.0 \pm
0.5$, leading to $\sigma_W = 0.36 \pm 0.03$. The small undulations in
the data on the high-$s$ side of the peak, in this example near
$s=1.5$ and again for larger $s$, is characteristic of poor data.
\label{Cupoor}
}
\end{figure}

\begin{figure}[t]
\caption[shrt]{
 Analysis of terrace width distributions of Pt(110) using
Wigner distributions.  The experimental points of Ref.~\cite{SBV}
are indicated by dots. As in Figs.~\ref{Cugood} and \ref{Cupoor}, the
short-dashed curve a conventional Gaussian fit, the long-dashed curve a
single-parameter Wigner fit, and the solid curve a two-parameter Wigner
fit.}
\label{PtSBV}
\end{figure}

\begin{figure}[t]
\caption[shrt]{
   The error in estimates $\tilde{A}_c$ of $\tilde{A}$ derived by
   using formulae for the mean and variance of the continuous
   generalized Wigner surmise TWD on discrete TWDs, for the physical range
of $\tilde{A}$.  $\langle L \rangle$ indicates the mean terrace
width in units of $a_{\perp}$.  For $\langle L \rangle = 2$ and 3, the ordinate
values have been divided by 1000 and by 50, respectively, to appear the
same vertical scale; evidently, discreteness for these narrow terraces
introduces unacceptably large errors, particularly as $\tilde{A}$
increases.  The smooth curves through these points, to guide the eye,
are dashed to distinguish them from the cases with broader terraces.
\label{discrete_rho}
}
\end{figure}

\begin{figure}[t]
\caption[shrt]{
   The error in estimates $\tilde{A}_c$ of $\tilde{A}$ derived by
   using formulae for the mean and variance of the continuous
   generalized Wigner surmise TWD on discrete TWD. The estimates evidently
    improve considerably with increasing
   $\langle L \rangle$ (broader terraces, with higher Miller indices).
\label{discrete_l0}
}
\end{figure}

\begin{figure}[t]
\caption[shrt]{
Standard deviations in fitted values of $\varrho$ due to statistical
fluctuations.  In each case, fits were made to terrace width distributions
consisting of $N_{\rm samp}\! = \! 801$ values of $s$
independently distributed according to a DGWD function.
Each circle represent a sampling over one million
uniformly-weighted two-parameter fits.
Each square represents a sampling over ten thousand
two-parameter fits, in which the weight for each point in the
TWD was weighted proportionally to $P(s)$.
Each diamond represents a sampling over ten thousand
applications of Eq.~(\protect\ref{eq:rho_of_sig2}).  The line
is the prediction of Eq.~(\protect\ref{e:rtvarrho}).
Both in terms of computational difficulty and statistical quality,
Eq.~(\protect\ref{eq:rho_of_sig2}) is clearly superior to
nonlinear least-squares fits.
\label{fig:sample_dRho}
}
\end{figure}

\begin{figure}[t]
\caption[shrt]{
  Twenty TWDs were simulated by drawing $N_{\rm samp}\! = \! 400$
values of $s$ according to a DGWD with $\varrho\! = \! 5$ and $\langle L
\rangle \! = \!
6$, indicated by the solid black
curve.  Each TWD was then fitted to a single-parameter CGWD, as in
Eqs.~(1)--(3). Shown are the TWDs with the smallest ($\bullet$) and
largest ($\Box$) values of $\chi^2$.  The fits to these are shown as
the black dashed curve and the $\times$'s, respectively.
\label{samplepics}
}
\end{figure}
\onecolumn
\setlength{\baselineskip}{4pt}
\noindent TABLE I: Tabulation of the results of fitting data from
various vicinal surfaces of copper to a Gaussian with three
parameters (labeled by subscript G) and to Wigner distributions with
one or with two adjustable parameters (labeled by subscripts 1 and 2,
respectively). The temperature in Kelvin is given in the first column
and the qualitative characterization (+ for good, 0 for fair, - for
poor) in the second.  The final column, labeled $\Delta
\mu$, indicates how much the mean (or first moment) computed directly
from the data exceeds the optimal mean obtained via the second
parameter in the two-parameter Wigner fit; using the notation
of Eq.~(\ref{e:Psa}), we have $\Delta \mu = \alpha^{-1} -1
\approx 1- \alpha$. Motivated by the discussion of Sec.~\ref{sec-sample},
we
include in the final column the standard deviation $\sigma_{\rm dir}$
evaluated
directly from the normalized (and adjusted to unit mean) histogram data.
\vspace{4mm}
\renewcommand{\arraystretch}{0.6}
\begin{tabular}{cc|cc|ccc|ccc|c|c} \hline
T&qual&$\sigma_G$&100$\chi^2_G$&$\varrho_1$&$\sigma_1$ & 100$\chi_1^2$ &
$\varrho_2$ &
$\sigma_2$ & 100$\chi^2_2$ & $\Delta \mu$ & $\sigma_{\rm dir}$\\
\hline \multicolumn{2}{l|}{(1,1,7)}&&&&&&&&& \\
298 & + &0.21 (1)&0.43&11.0 (6)&0.21 (1)&0.32&11.0 (6)&0.21 (1)&0.35&-0.00
(1) & 0.23 \\ \hline
\multicolumn{2}{l|}{(1,1,13)}&&&&&&&&&& \\
285 & + &0.19 (2)&1.31&9.7 (12)&0.22 (1)&2.21&10.0 (8)&0.21 (1)&1.16&0.04
(1) & 0.25 \\
300 & + &0.25 (1)&0.73&6.4 (5) &0.26 (1)&0.78& 6.5 (3)&0.26 (1)&0.36&0.04
(1) & 0.28 \\
303 & 0 &0.26 (1)&0.58&5.7 (6) &0.28 (1)&1.00& 5.9 (3)&0.27 (1)&0.31&0.05
(1) & 0.36 \\
320 & + &0.27 (2)&0.61&5.2 (4) &0.29 (1)&0.79& 5.3 (4)&0.29 (1)&0.39&0.04
(1) & 0.33 \\
326 & - &0.27 (2)&1.52&2.8 (6) &0.38 (3)&3.34& 3.5 (4)&0.34 (1)&1.44&0.13
(3) & 0.45 \\
330 &-/0&0.28 (2)&1.86&3.9 (6) &0.33 (1)&1.80& 4.2 (4)&0.32 (2)&1.16&0.06
(2) & 0.35 \\
338 & + &0.25 (1)&0.52&5.3 (6) &0.29 (1)&1.13& 5.6 (3)&0.28 (1)&0.31&0.06
(1) & 0.34 \\
348 & + &0.27 (1)&0.88&4.4 (5) &0.31 (1)&1.59& 4.8 (4)&0.30 (1)&0.57&0.07
(1) & 0.36 \\
350 &0/+&0.27 (1)&0.16&5.0 (4) &0.29 (2)&0.89& 5.1 (3)&0.29 (1)&0.23&0.06
(1) & 0.40 \\
358 & 0 &0.21 (2)&0.79&5.6 (10)&0.28 (3)&2.71& 6.8 (6)&0.26 (1)&0.87&0.09
(1) & 0.38 \\
368 & - &0.25 (2)&1.77&3.1 (5) &0.36 (3)&2.89& 3.9 (4)&0.33 (1)&1.57&0.12
(2) & 0.45 \\
378 & - &0.30 (4)&3.52&2.5 (7) &0.39 (3)&4.20& 3.0 (5)&0.36 (3)&2.48&0.14
(3) & 0.46 \\
\hline \multicolumn{2}{l|}{(1,1,19)}&&&&&&&&&& \\
290 & - &0.40 (4)&3.74&2.7 (4)&0.38 (2)&2.33&2.7 (5)&0.38 (2)&2.49&-0.01
(3) & 0.34 \\
300 &-/0&0.24 (2)&2.94&3.1 (6)&0.36 (4)&5.09&4.1 (6)&0.32 (2)&2.69& 0.12
(3) & 0.39 \\
308 & 0 &0.31 (1)&0.71&4.4 (3)&0.31 (1)&0.32&4.3 (2)&0.31 (1)&0.30& 0.01
(1) & 0.30 \\
320 & + &0.25 (1)&0.54&6.7 (3)&0.26 (1)&0.36&6.7 (2)&0.26 (1)&0.23& 0.02
(1) & 0.27 \\
360 & + &0.27 (2)&1.74&5.7 (5)&0.28 (1)&0.84&5.8 (3)&0.28 (1)&0.64& 0.03
(1) & 0.29 \\
370 & - &0.30 (3)&4.20&4.0 (7)&0.32 (3)&3.30&4.3 (6)&0.31 (1)&2.71& 0.06
(2) & 0.31 \\
\hline \multicolumn{2}{l|}{(11,7,7)}&&&&&&& && & \\
296 &+&0.26 (2)&0.55&5.7 (4)&0.28 (1)&0.58&5.8 (3)&0.28 (1)&0.27&0.04 (1)
& 0.30 \\
301 &+&0.27 (2)&0.80&6.0 (4)&0.27 (1)&0.47&6.0 (5)&0.27 (1)&0.46&0.01 (2)
& 0.28 \\
306 &+&0.28 (2)&0.36&4.8 (3)&0.30 (2)&0.48&4.9 (2)&0.30 (1)&0.25&0.04 (1)
& 0.36 \\
323 &+&0.29 (1)&0.20&5.0 (1)&0.29 (1)&0.10&5.0 (2)&0.29 (1)&0.10&0.00 (1)
& 0.31 \\
\hline \multicolumn{2}{l|}{(19,17,17)}&&&&&&&&&  & \\
305 & - &0.23 (2)&2.77&5.3 (7)&0.29 (2)&3.62&6.3 (7)&0.27 (2)&2.32&0.08
(1) & 0.33 \\
313 & - &0.25 (2)&2.29&4.3 (5)&0.32 (1)&3.15&5.0 (4)&0.29 (2)&1.90&0.09
(1) & 0.38 \\
333 &-/0&0.34 (2)&1.19&3.3 (3)&0.35 (1)&0.90&3.3 (3)&0.35 (1)&0.82&0.03
(2) & 0.37 \\
353 &-/0&0.31 (2)&1.11&4.0 (3)&0.32 (1)&0.79&4.1 (2)&0.32 (1)&0.71&0.03
(2) & 0.33 \\
  \hline \multicolumn{2}{l|}{(23,21,21)}&&&&&&&&&  & \\
318 &-/0&0.24 (1)&0.92&7.1 (4)&0.25 (2)&1.10&7.2 (4)&0.25 (1)&0.74& 0.04
(1) & 0.33 \\
328 & 0 &0.29 (1)&1.06&5.3 (3)&0.29 (1)&1.05&5.4 (3)&0.29 (1)&1.01&-0.01
(1) & 0.31
\end{tabular}
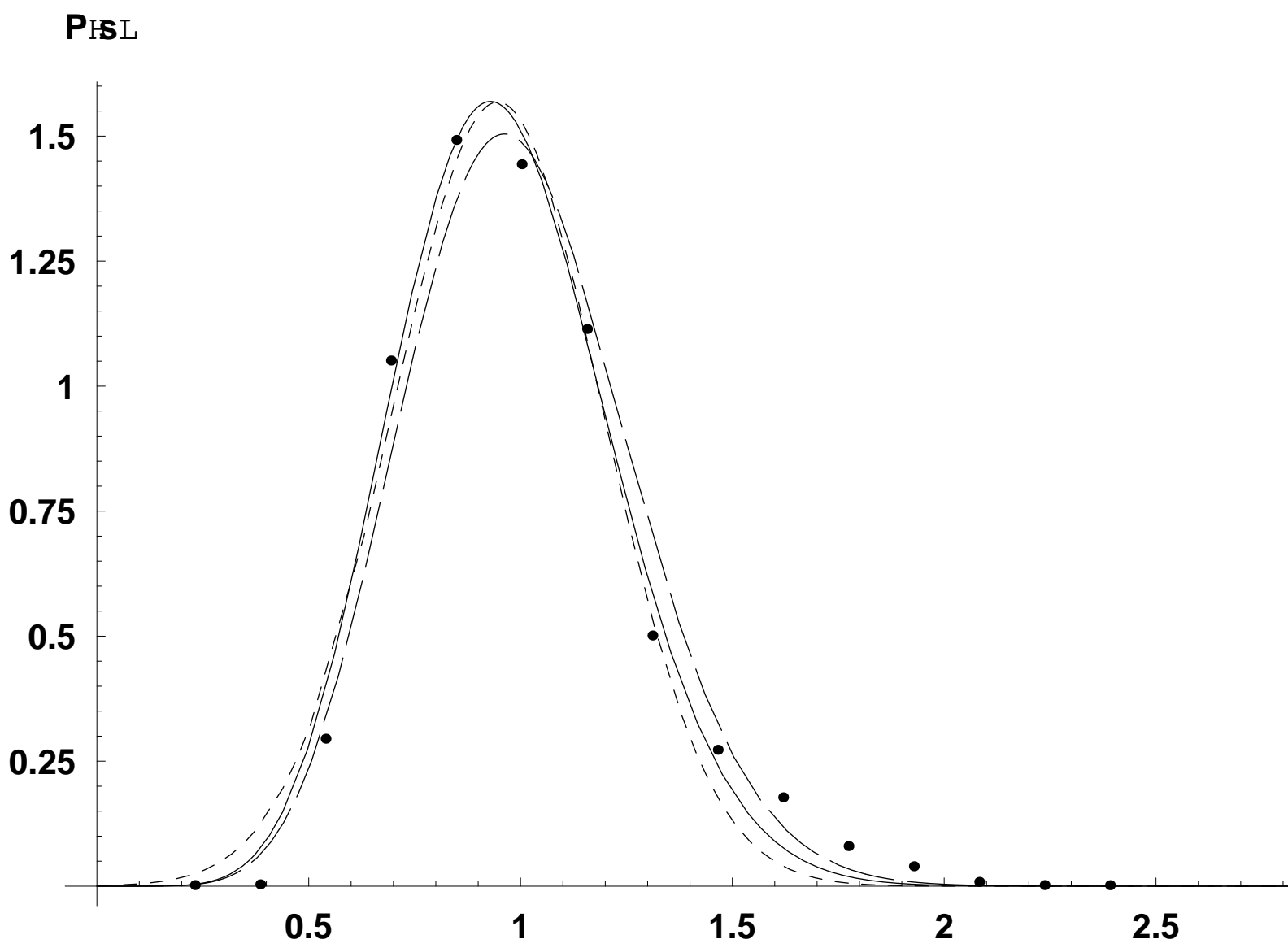
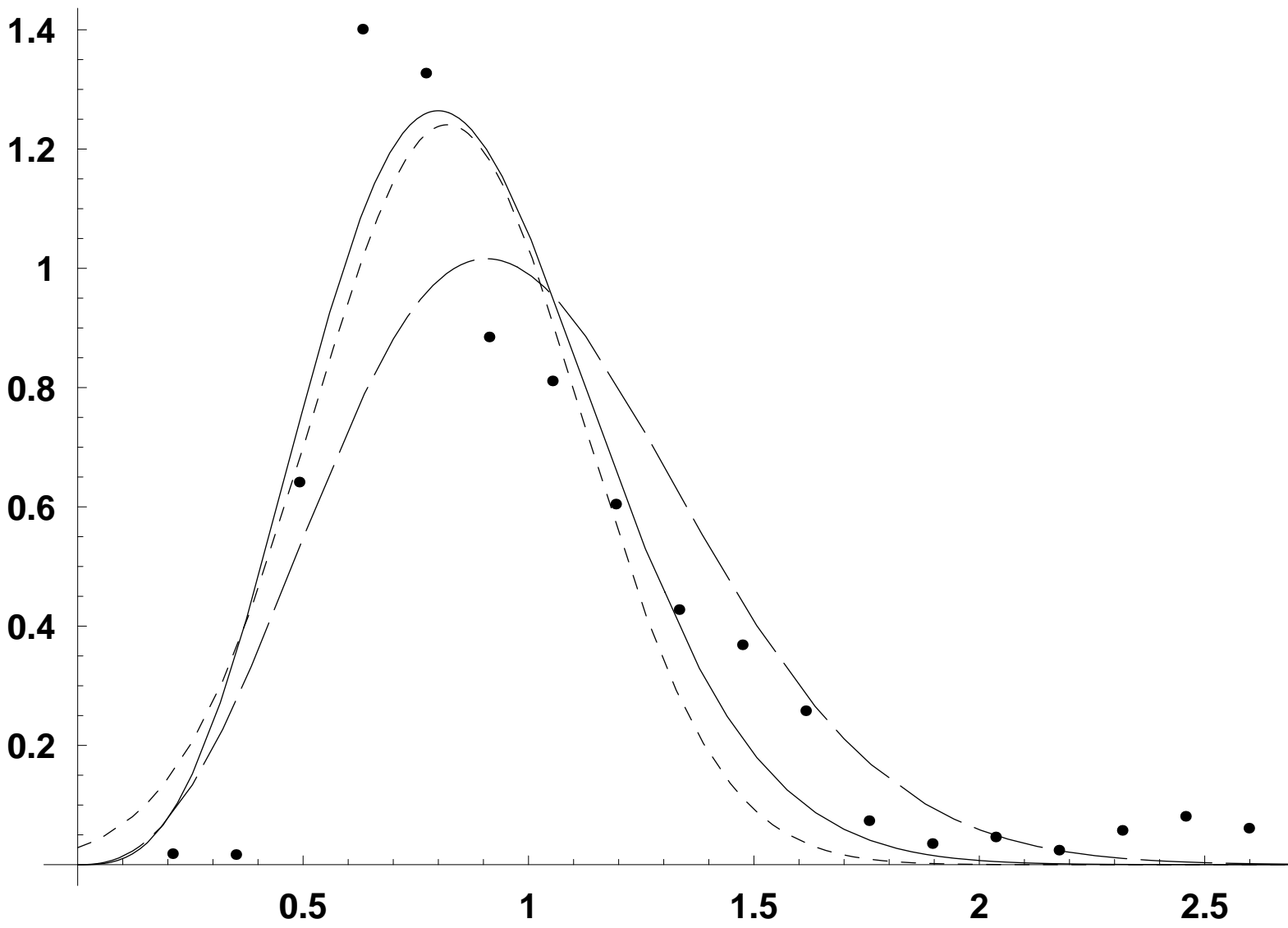
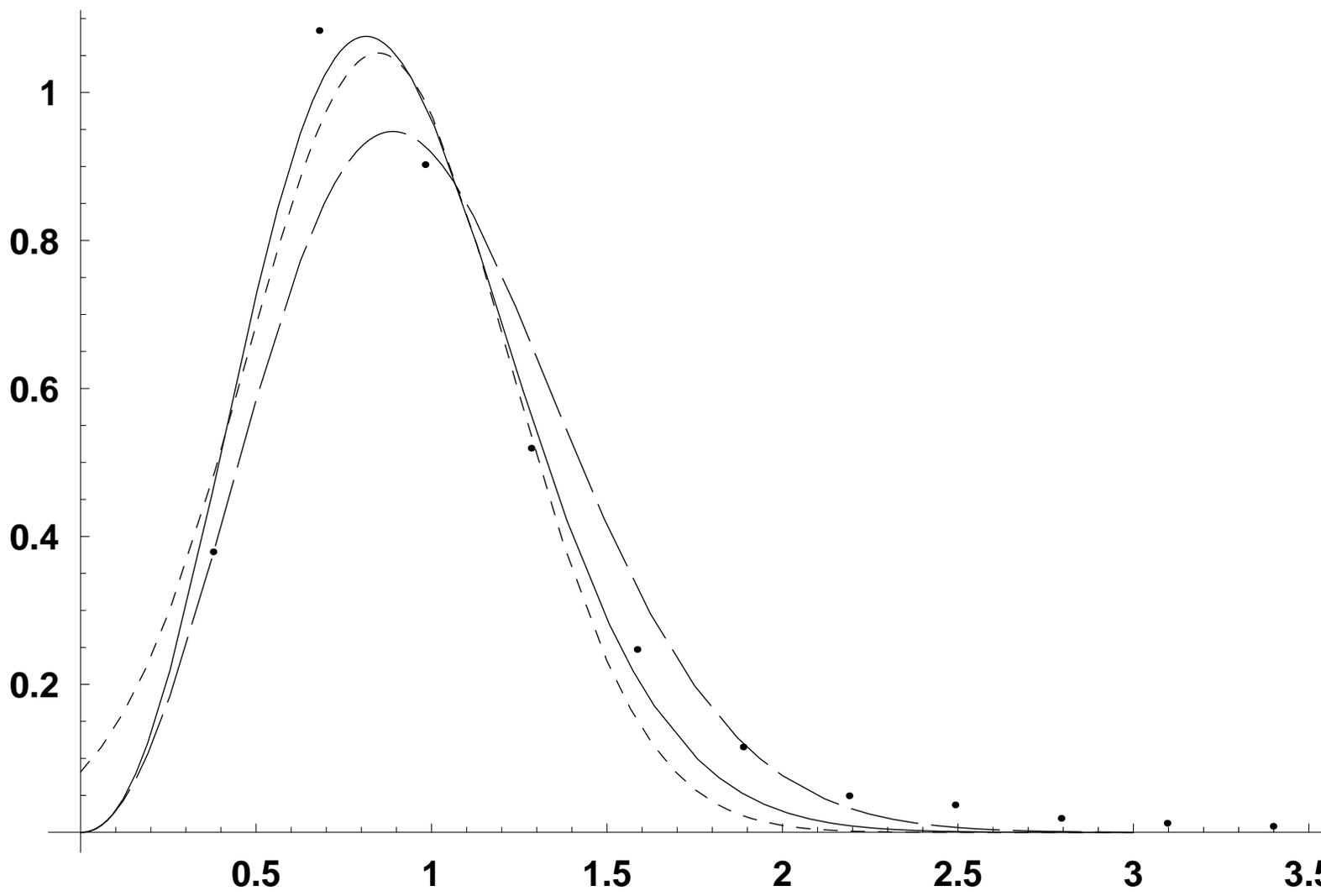
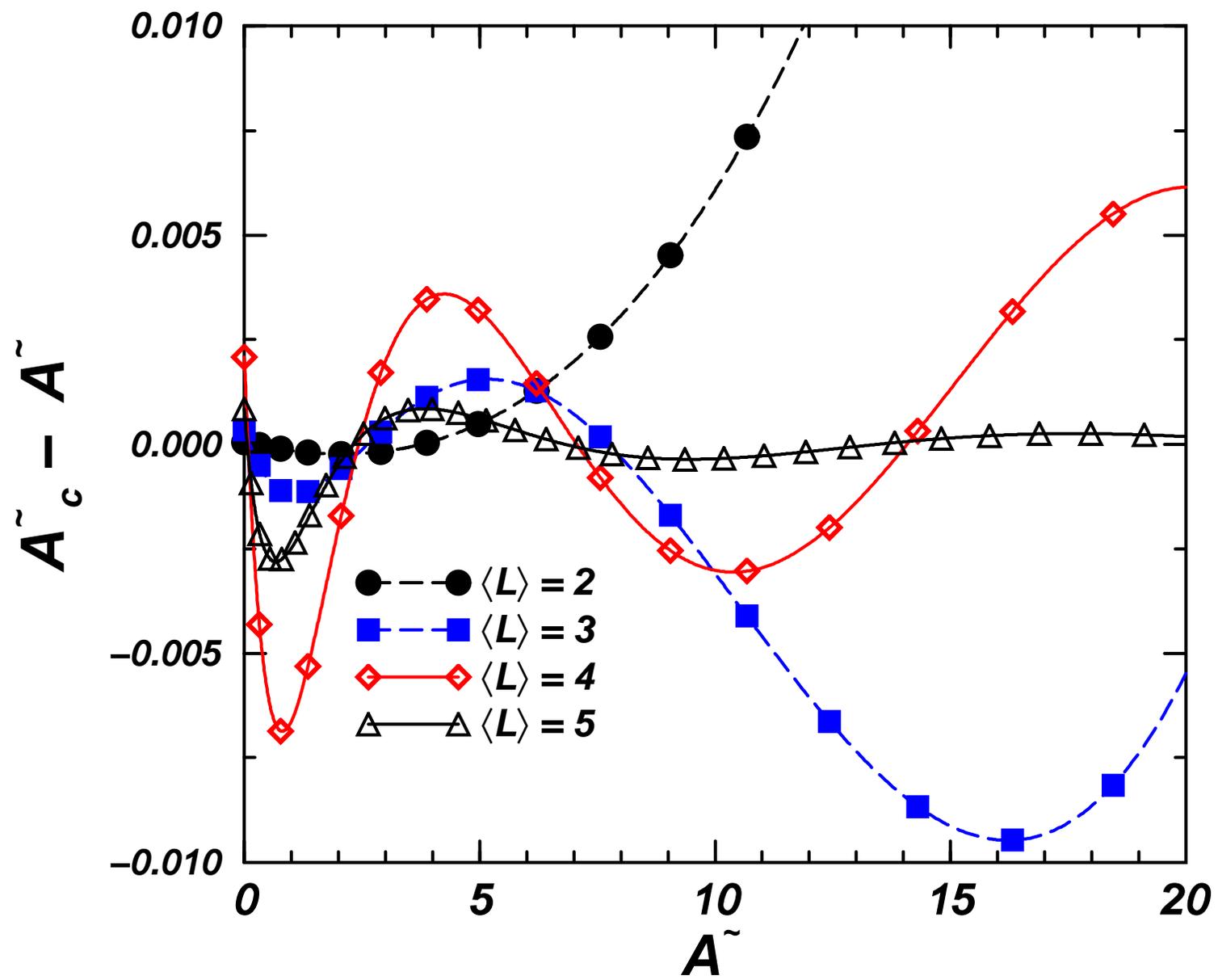
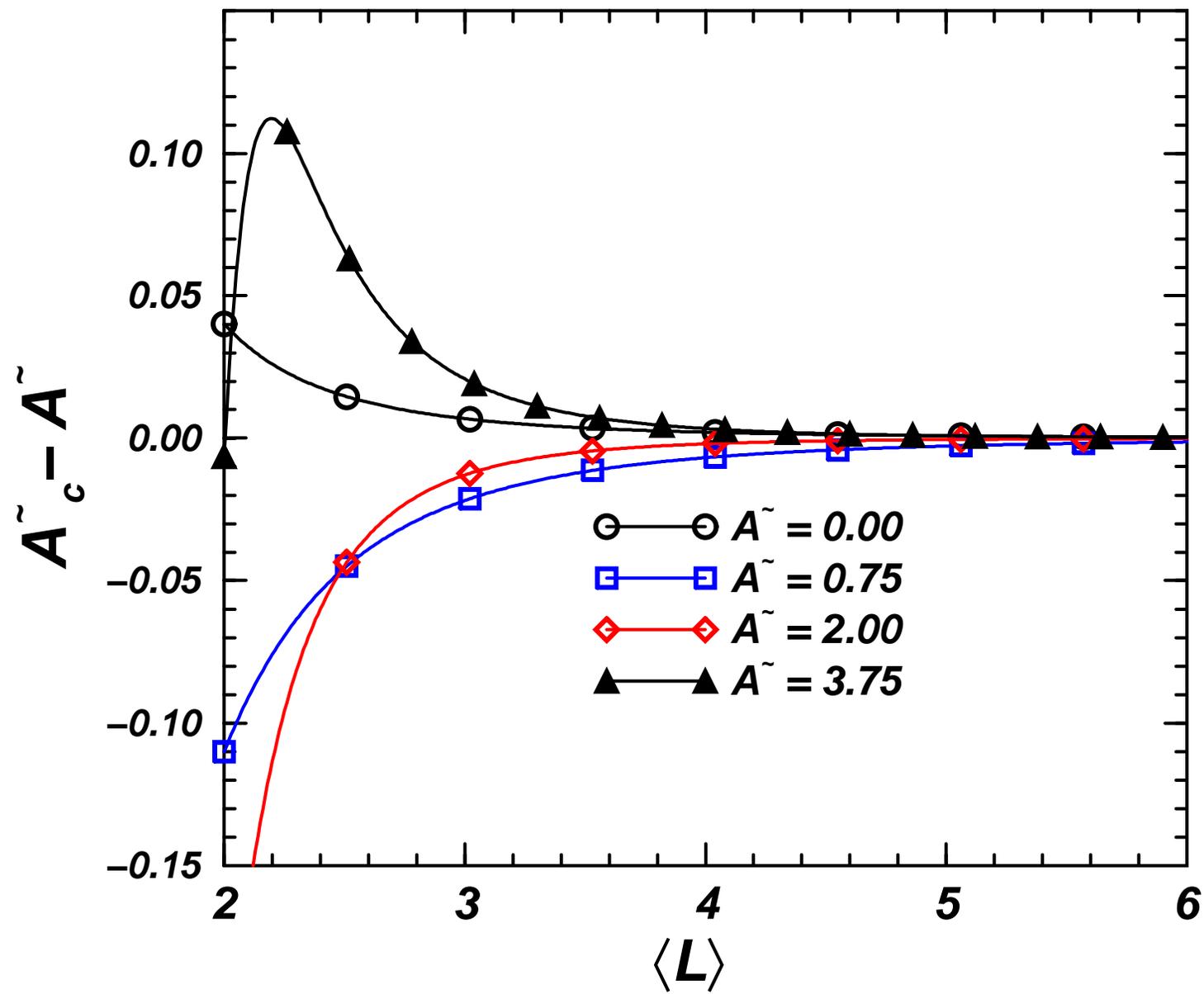
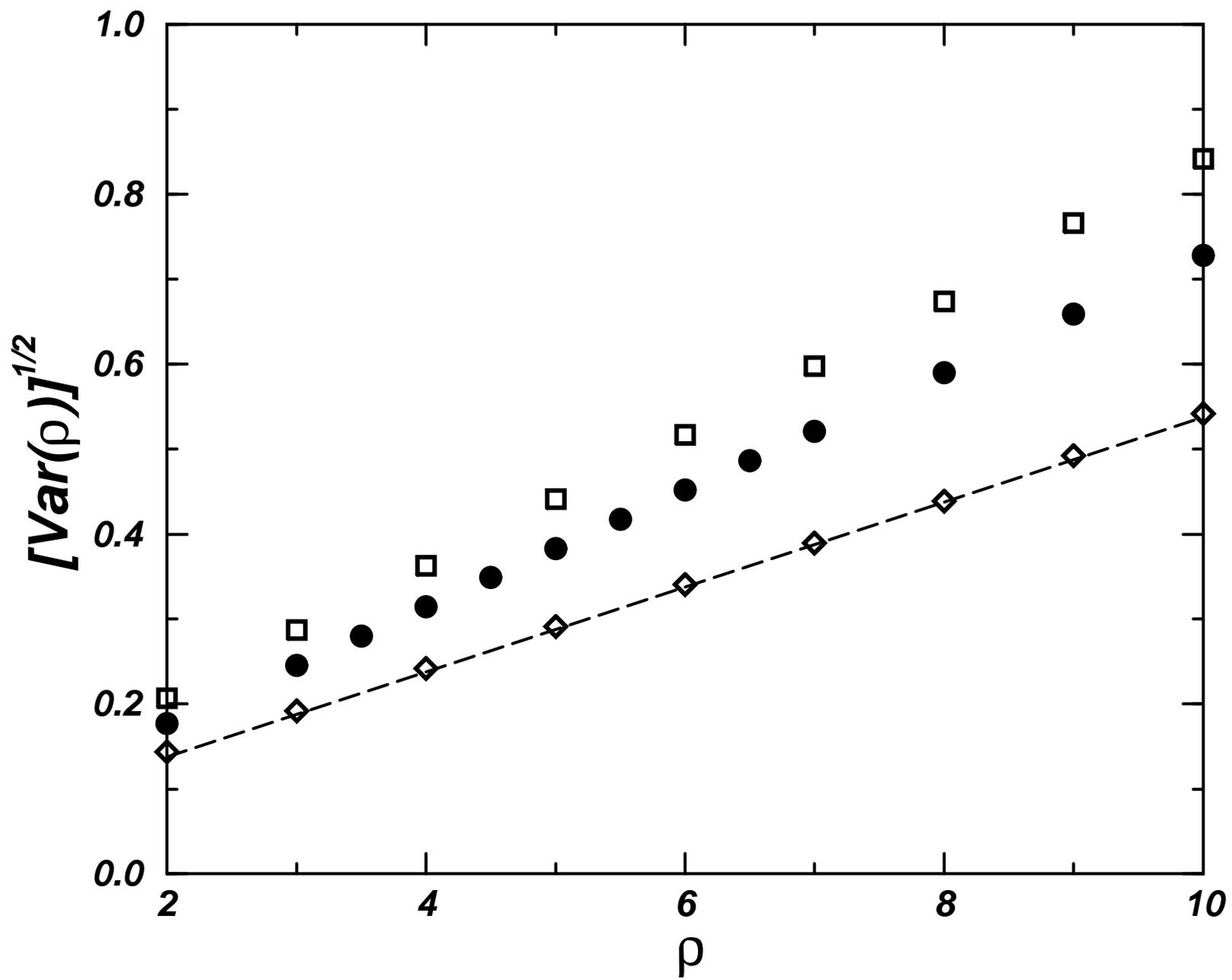
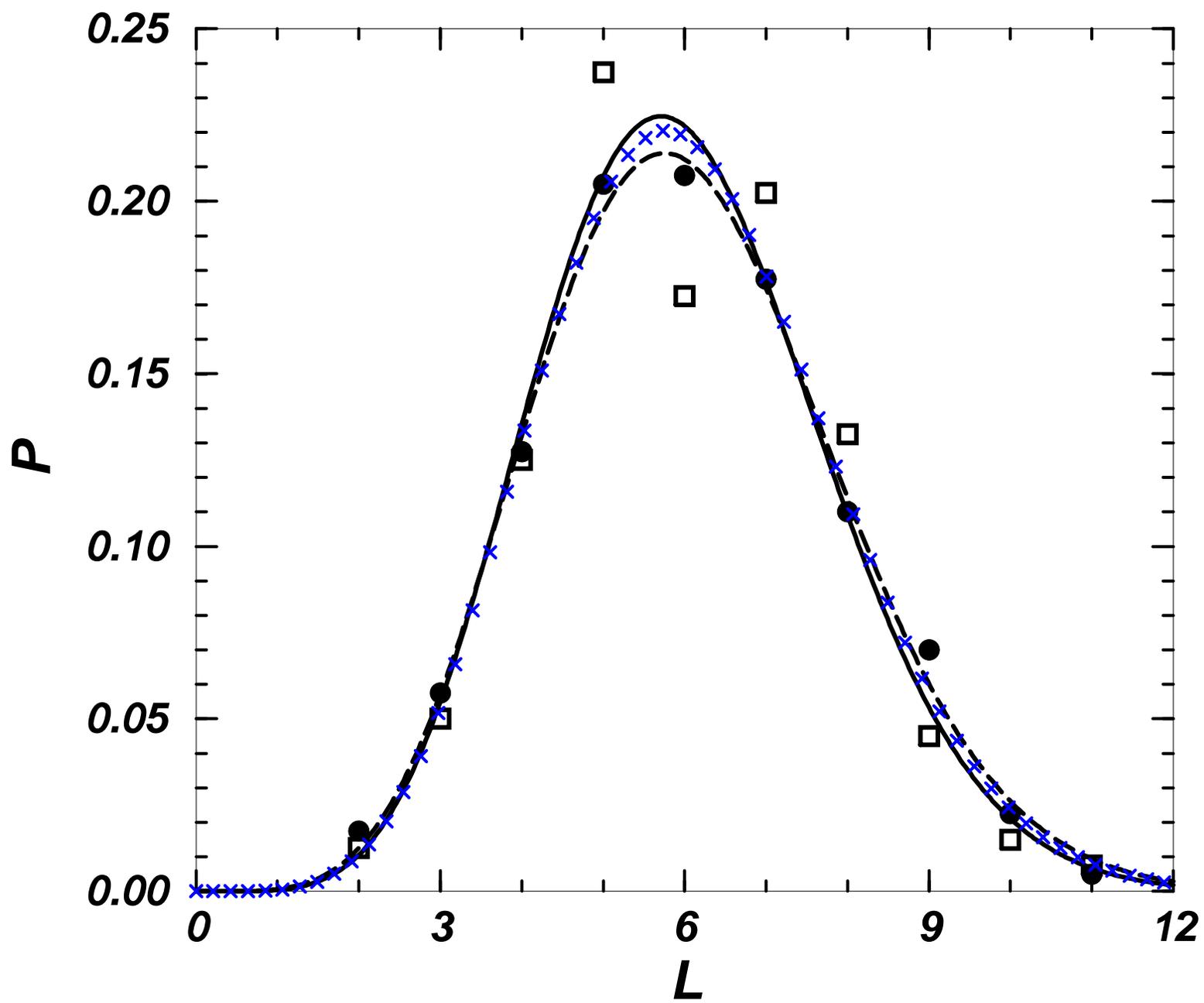
\end{document}